\documentclass[useAMS,usenatbib]{mn2e}
\usepackage{graphicx}
\usepackage{epstopdf}
\usepackage{amsmath}
\title[]{Bondi accretion in the finite luminous region of elliptical galaxies }
\author[M. Samadi; S. Zanganeh; S. Abbassi ]{
 M. Samadi$^{1,2}$\thanks{$samadimojarad$@um.ac.ir}; S. Zanganeh$^{1}$  \thanks{$zanganeh.sahar$@um.ac.ir}; S. Abbassi$^{1}$ \thanks{$abbassi$@um.ac.ir}
   \\
$^{1}$Department of Physics, Faculty of Science, Ferdowsi University of Mashhad, Mashhad, 91775-1436, Iran\\
$^{2}$Research Institute for Astronomy and Astrophysics of Maragha (RIAAM)- Maragha, P. O. Box: 55134 - 441, Iran}

\date{}
\begin{document}
\pagerange{\pageref{firstpage}--\pageref{lastpage}} \pubyear{2019}

\maketitle \label{firstpage}

\begin{abstract}
The classical Bondi model is adopted to study accretion onto the finite luminous region around the central massive black hole (MBH) in an elliptical galaxy. Unlike Bondi (1952), we define the boundary conditions at a certain finite radius ($r_f$) instead of at the infinity and examine the variation of solutions for a simple case. In the following, we consider the special case of a MBH at the center of a Hernquist galaxy and involve the gravity and luminosity of its own galaxy.
Our results in the first part show that kinitic energy at the final radius is ignorable even for not so far away from the center. Moreover, the mass accretion rate will be approximately equal to its Bondi value if the final radius ($r_f$) becomes about 2-3 orders of magnitude larger than semi-Bondi radius, i.e. $GM/c_{sf}^2$ (where $M$ and $c_{sf}$ are the mass of the central object and the sound speed at $r_f$). In the second part, adding the two extra forces of gravity and radiation in the momentum equation let us know that the maximum possible of accretion rate increases with greater characteristic linear density of galaxy and lower radiation. 
\end{abstract}


\section{Introduction}
Nowadays, we have access to some fundamental models for mathematical interpretation of very active astronomical events at the universe. The most powerful source of energy in such events arises from the weakest force among four types of forces, i.e. gravity. This weak force plays a vital role for providing energy in black holes, neutron stars and other compact objects. On the other hand, in protostars when the nuclear energy is not still available, the required energy for making them visible in the heavens is provided by making use of their gravitational binding energy through contraction (Kato et al. 2008). It might be right that the first simplest model in this framework was supposed by Bondi (1952). The Bondi accretion model describes a steadily inflowing current of an adiabatic gas just at the radial direction toward a central mass point. In fact, this simple model is sometimes the only available tool for the first step of some studies in numerical simulations when that the resolution of observations is not sufficient in determination of all properties of infalling gas very far from the center. Based on the Bondi accretion, we are now able to estimate the mass accretion rate on massive black holes (MBH) at the center of galaxies, with observational data of the gas density, temperature in the vicinity of the MBH and also the sonic radius (Loewenstein et al. 2001; Pellegrini 2005, 2010; Allen et al. 2006;  Sijacki et al. 2007; Di Matteo et al. 2008; Gallo et al. 2010; Barai et al. 2011, 2012; McNamara et al. 2011; Wong et al . 2014; Russell et al. 2015; Park et al. 2017; Beckmann et al.
2018).

Since 1952, some researches have been done to involve more details in the spherical accretion model of Bondi (Begelman 1978, 1979; Ramirez-Velasquez et al. 2019, Yalinewich et al. 2018). Regarding radiation in this problem, several attempts have been made to investigate the effects of radiation on the radial current of gas toward the center of a spherical system. Maraschi et al. (1978) considered an accreting source with a radiation power in the order of the Eddington limit. They took into account the radiation pressure due to Thomson scattering and neglected the gas pressure. Begelman (1978) and (1979) concentrated on the spherical accretion onto black holes with dominant pressure of radiation and concluded  that the diffusive luminosity at infinity cannot exceed $0.6L_{Edd}$. Fukue (2001) examined the radiation effects in the spherical accretion of an ionized gas onto a luminous central object. He could obtained an approximate relation for the accretion rate and proved the mass accretion rate reduces and depends on the luminosity of the central gravitating mass. Mattews \& Guo (2012) presented a research to compare  a steady accretion of a radiating gas with the classic Bondi accretion. Recently, the Bondi accretion with extra effects of radiation and gravity of stars has been applied for early-type galaxies by Korol et. al (2016). They calculated the deviation from the true values of the estimated Bondi radius and mass accretion rate due to adopting as boundary values for the density and temperature those at a finite distance from the central mass. 
Similar projects have been carried out by Ciotti \& Pellegrini (2017), (2018) for spherical accretion in some galaxies with a central black hole. In the first paper, they found out the whole accretion solution is reachable in an analytical way for accretion problem in Jaffe (1983) and Hernquist galaxy models. They focused on the case of isothermal accretion in galaxies with central MBHs and with radiation pressure contributed by electron scattering in the optically thin regime. They could obtained the radial profile of the Mach number by using a known mathematical function and proved that the value of the critical accretion parameter can be analytically calculated for the two cases of Jaffe and Hernquist galaxies. On the other hand, the Bondi accretion has been applied for two- component Jaffe (1983) galaxy models with a centeral MBH by Ciotti \& Ziaee Lorzad
(2018). In the second paper of Ciotti \& Pellegrini, this approach has been followed for studying the extra influence of radiation pressure contributed by electron scattering in the optically thin regime.

 In this paper, we aim to revisit spherical Bondi problem with setting the boundary conditions at a finite radius instead of infinity. Our second aim is to study the influence of the luminosity of galaxy and from the accretion process. We especially concentrate on a Hernquist galaxy and employ its proposed mathematical model for providing two extra forces both in the radial direction, firstly the attraction force of gravity, and secondly repulsion force exerted from the galaxy's radiation. In the next section, the basic equations will be presented.

\section{Basic Equations}
In the classical Bondi, an infinite cloud of gas surrounding a star of mass $M$. Due to the gravity of the star, the gas at infinity (with uniform density $\rho_\infty$ and pressure $p_\infty$) begins moving along the radial direction so the accretion occures in the spherical symmetry and steady pattern. The density ($\rho$) and pressure ($p$) of the gas are assumed to obey the polytropic relation, 
\begin{equation}
p=K\rho^\gamma
\end{equation}
where $K$ and $\gamma$ are a constant and the ratio of specific heats, respectively. Two other constants (i.e. mass accretion rate and Bernoulli number) are obtained from the continuity and momentum equations, 
\begin{displaymath}
\frac{\partial\rho}{\partial t}+\nabla\cdot (\rho\textbf{v})=0,
\end{displaymath}
\begin{displaymath}
\frac{D\textbf{v}}{Dt}=-\frac{\nabla p}{\rho}-\nabla \Phi 
\end{displaymath}
where $\textbf{v}$ is velocity vector and $\Phi$ is the gravitational potential. Moreover, the Lagrangian derivative, $D/Dt$ is defined by $\partial/\partial t+\textbf{v}\cdot\nabla$. For a steady state ($\partial/\partial t\rightarrow 0$) and spherical symmetric ($\nabla\rightarrow d/dr$) flow those two equations simplified to 
\begin{equation}
\frac{d}{dr}(r^2\rho v_r)=0,\hspace*{0.2cm}\rightarrow r^2\rho v_r=\frac{\dot{M}}{4\pi}=cte.
\end{equation}
\begin{equation}
 v_r\frac{dv_r}{dr} = -\frac{1}{\rho}\frac{dp}{dr}-\frac{d\Phi}{dr},
\end{equation}
where $\dot{M}$ is the mass accretion rate. If we suppose that the gravitational potential is just provided by the central object of the cloud and the gravity of particles in the cloud itself is ignorable, then we have $\Phi=-GM/r$. The integral of Eq. (3) from a certain distance, $r_f$, to the radius, $r$ leads us to find
\begin{displaymath}
\frac{v^2-v_f^2}{2}-GM\big(\frac{1}{r}-\frac{1}{r_f}\big)=-\int_{r_f}^r \frac{1}{\rho}\frac{dp}{dr}dr
\end{displaymath}
the term in the right hand side of this equation becomes equal to 
\begin{displaymath}
\int_{r_f}^r \big(\frac{1}{\rho}\frac{dp}{dr}\big)dr=\int_{\rho_f}^\rho \big(\frac{dp}{d\rho}\big)\frac{d\rho}{\rho}
\end{displaymath}
where $\rho_f$ is the density at $r_f$. Employing Eq.(1) in it, we have 
\begin{displaymath}
...=\int_{\rho_f}^\rho (K\gamma \rho^{\gamma-2})d\rho=K\gamma\frac{\rho^{\gamma-1}-\rho_f^{\gamma-1}}{{\gamma-1}}
\end{displaymath}
Regarding the sound speed defintion $c_s^2=dp/d\rho=K\gamma\rho^{\gamma-1}$, this term is simplified as
\begin{equation}
\int_{r_f}^r \frac{1}{\rho}\frac{dp}{dr}dr=\frac{c_s^2-c_{sf}^2}{\gamma-1}
\end{equation}
Substituting this result in Eq.(3) and rearrange the terms, we have:
\begin{equation}
\frac{v^2}{2}-\frac{GM}{r}+\frac{c_s^2}{\gamma-1}=\frac{v_f^2}{2}-\frac{GM}{r_f}+\frac{c_{sf}^2}{\gamma-1}
\end{equation}
For solving this equation, we should employ dimensionless quantities as,
\begin{equation} 
\begin{cases} 
r=R r_e,\\
v=V c_{sf},\\ 
c_s=C_s c_{sf}\\
\rho=Z \rho_f 
\end{cases} 
\end{equation} 
where $r_e= GM/c_{sf}^2$. Then Eq.(4) takes the non-dimensional form as:
\begin{equation}
\frac{V^2}{2}-\frac{1}{R}+\frac{C_s^2}{\gamma-1}=\frac{V_f^2}{2}-\frac{1}{R_f}+\frac{1}{\gamma-1}
\end{equation}
There are two unknowns, $V$ and $C_s$ which should be found with respect to $R$. With using  dimensionless quantities in Eq.(5), Eq.(2) also changes as,
\begin{equation}
R^2 V Z= \lambda,
\end{equation}
where $\lambda$ is the dimensionless accretion parameter and has been found by using $\dot{M}_e$ as:
\begin{displaymath}
\lambda=\frac{\dot{M}}{\dot{M}_e},\hspace*{0.5cm}\dot{M}_e=4\pi \frac{G^2M^2}{c_{sf}^3}\rho_f
\end{displaymath}
Here we need to replace $C_s^2$ with a term including $Z$, so we use the relationship of $c_s^2=K\gamma\rho^{\gamma-1}$ and write it for two radii of $r$ and $r_f$, so we will have
\begin{displaymath}
\frac{c_s^2}{c_{sf}^2}=\big(\frac{\rho}{\rho_f}\big)^{\gamma-1},
\end{displaymath}
So we use non-dimensional density of $Z$ instead of $C_s$ by the following relationship,
\begin{equation}
C_s^2=Z^{\gamma-1},
\end{equation}
To solve these algebraic equations, it requires to use Mach number, $u$: 
\begin{equation}
u=\frac{V}{C_s}=VZ^{-(\gamma-1)/2},
\end{equation}
\begin{figure*} 
\centering 
\includegraphics[width=190mm]{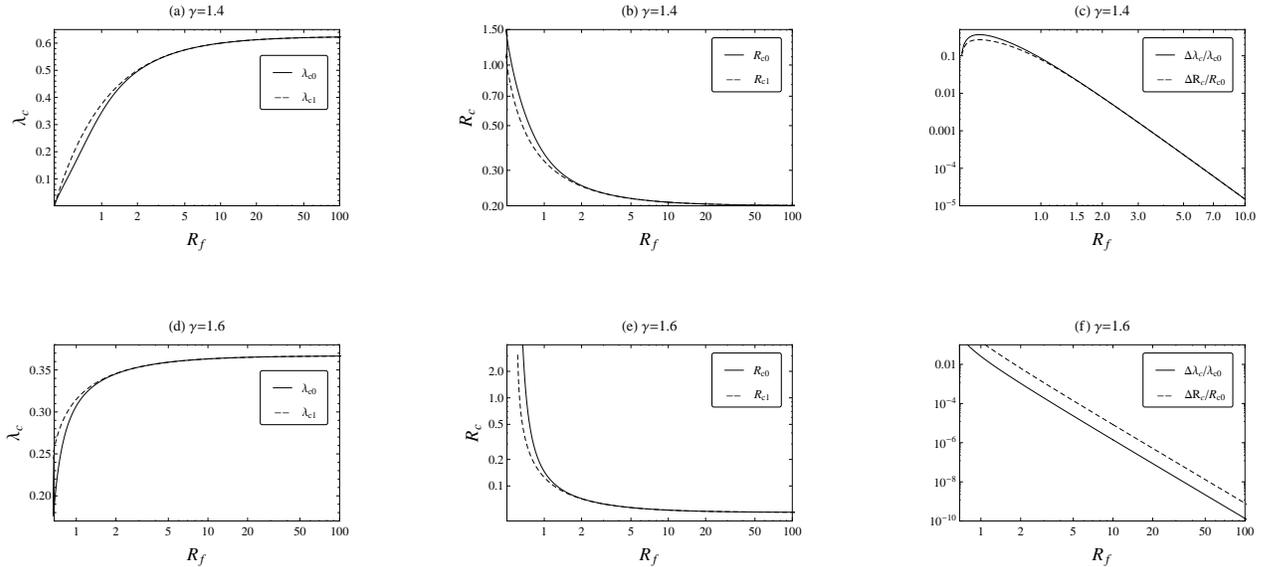} 
\caption{Variation of critical dimensionless mass accretion rate, $\lambda_c$, critical radius, $R_c$, and their error percentages, $\Delta\lambda/\lambda_{c0}$, $\Delta R_c/R_{c0}$, with respect to the final radius of a certain region of accretion, $R_f$. These quantities have been obtained in two ways of approximate and exact solutions. The dashed lines are related to exact solutions that we have included the kinetic energy of particles at the end of accretion region. } 
\end{figure*} 
Now we can find the following relationships for $Z$, $C_s^2$ and $V^2$
\begin{equation}
\begin{cases}
Z=\big(\frac{\lambda}{R^2 u}\big)^{2/(\gamma+1)},\\
C_s=\big(\frac{\lambda}{R^2 u}\big)^{(\gamma-1)/(\gamma+1)},\\
V=u\big(\frac{\lambda}{R^2 u}\big)^{(\gamma-1)/(\gamma+1)},
\end{cases}
\end{equation}
Then Eq.(7) becomes
\begin{equation}
\big(\frac{u^2}{2}+\frac{1}{\gamma-1}\big)\big(\frac{\lambda}{R^2 u}\big)^{2\frac{\gamma-1}{\gamma+1}}=\frac{1}{R}+\frac{1}{\gamma-1}+\frac{V_f^2}{2}-\frac{1}{R_f}
\end{equation}
Now it would be better to separate the terms in two groups, one forming $f(u)$ and the other providing $g(R)$, as:
\begin{equation}
f(u)= u^{-2\frac{\gamma-1}{\gamma+1}}\bigg(\frac{u^2}{2}+\frac{1}{\gamma-1}\bigg)
\end{equation}
\begin{equation}
g(R)=R^{4\frac{\gamma-1}{\gamma+1}}\bigg(\frac{1}{R}+\frac{1}{\gamma-1}+\frac{V_f^2}{2}-\frac{1}{R_f}\bigg)
\end{equation}
All of our equations are similar to ones of Bondi (1952), just the only difference is related to the final radius, ($r_f$) in the integration of Eq.(3). Bondi solved the problem with $r_f=\infty$, hence both kinetic ($KE_f$) and potential ($PE_f$) energies were assumed to be zero. Here, we have two extra terms in the function of $g(R)$, both of them are apparently known. Nevertheless, the two terms are dependent together with Eq.(7), means
\begin{displaymath}
V_f =\frac{\lambda}{R_f^2}
\end{displaymath}
where we have used $Z_f=1$. As seen, the velocity of particles in the final radius depends on the mass accretion parameter, $\lambda$, thus $g(R)$ is not only the function of $R$. Now calculating the extreme value of $g(R)$ gives us the critical radius of $R_c$ as bellow, 
\begin{equation}
R_c=\frac{5-3\gamma}{4[1+A(\gamma-1)]}
\end{equation}
where $A$ is the summation of two extra terms at the final radius, ($A=KE_f+PE_f$). We refer to the other function of this problem, $f(u)$ which is independent of the final point of accretion process, hence its extreme value does not change comparing with Bondi's work and remains unity. The most important quantity here is the maximum possible value of the mass accretion rate parameter. Generally for finding $\lambda$, firstly we should know $g(R)$ and $f(u)$, then use the following calculation,
\begin{equation}
\lambda=\bigg[\frac{g(R)}{f(u)}\bigg] ^ {\frac{\gamma+1}{2(\gamma-1)}}
\end{equation}
However, as we mentioned $g(R)$ is implicitly dependent on $\lambda$, so we suggest two ways to estimate $\lambda$. The simple way is to neglect $KE_f$ and the other way is to solve exactly this problem with $g(R,\lambda)$. Therefore, we can have two groups of critical values, $R_c$ and $\lambda_c$.   
The first group with $A=-1/R_f$,
\begin{equation}
R_{c0}=\frac{5-3\gamma}{4[1-(\gamma-1)/R_f]}
\end{equation}
\begin{equation}
\lambda_{c}=\bigg[\frac{g(R=R_{c})}{f(u=1)}\bigg] ^ {\frac{\gamma+1}{2(\gamma-1)}}
\end{equation}
Using Eq.(13) and (16) in Eq.(17) we have,
\begin{equation}
\lambda_{c0}=\bigg(\frac{1}{2}\bigg)^{\frac{\gamma+1}{2(\gamma-1)}}\bigg\{\frac{5-3\gamma}{4[1-(\gamma-1)/R_f]}\bigg\}^{\frac{3\gamma-5}{2(\gamma-1)}},
\end{equation}
The second way that leads us to find the exact critical value of the accretion parameter is to solve the following algebraic equation,
\begin{displaymath}
\frac{\gamma+1}{\gamma-1}\big(\frac{\lambda_{c1}}{R^2}\big)^{2\frac{\gamma-1}{\gamma+1}}-\bigg(\frac{\lambda_{c1}}{R_f^2}\bigg)^2=2\bigg(\frac{R_f-R}{RR_f}+\frac{1}{\gamma-1}\bigg)
\end{displaymath}
\begin{figure*} 
\centering 
\includegraphics[width=190mm]{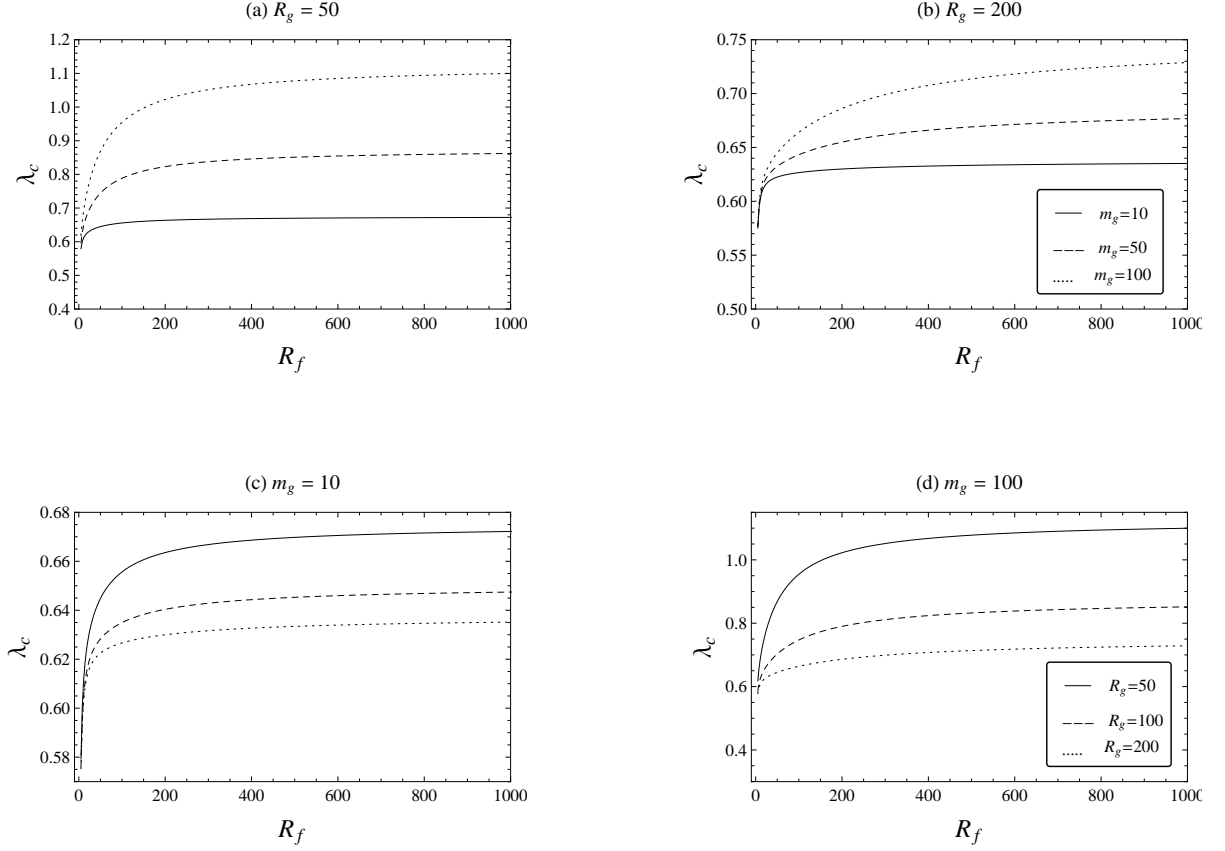} 
\caption{Variation of critical radius, $R_c$ with respect to the final radius of a certain region of accretion, $R_f$. Here the polytropic index, $\gamma$ is equal to 1.4 and we have separately examined the effects of $m_g(=M_{galaxy}/M_{BH})$ and $R_g(=r_g/r_e, r_g$ is the characteristic scale-length of galaxy, $r_e=GM_{BH}/c_{sf}^2$). Notice that for these plots, we have neglected the kinetic energy of particles at the end of selected accretion region. } 
\end{figure*} 
\begin{figure*} 
\centering 
\includegraphics[width=190mm]{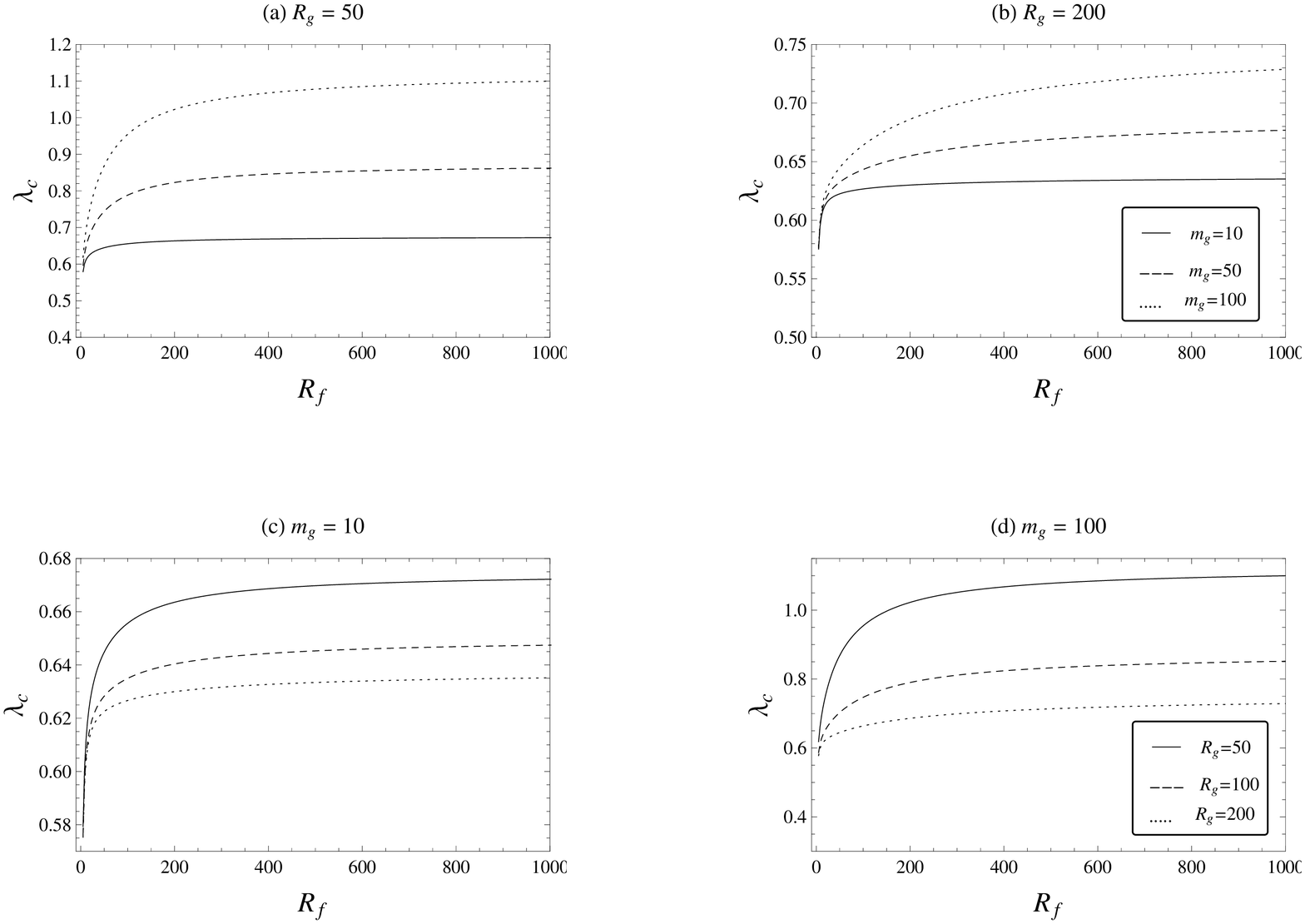} 
\caption{Variation of critical accretion parameter, $\lambda_c$ with respect to the final radius of a certain region of accretion, $R_f$. Here the polytropic index, $\gamma$ is equal to 1.4 and we have examined the effects of $m_g$ and $R_g$, separately. Notice that for these plots, we have assumed $V_f\sim 0$, means that particles on the sphere of $R_f$ are approximately at rest.} 
\end{figure*} 

The results of these two ways are shown in figure (1). It is interesting that just for distances smaller than about $2r_e$, the difference between $\lambda_{c0}$ and $\lambda_{c1}$ is recognizable for both $\gamma=1.4$ and $\gamma=1.6$. The critical radius, $R_c$ also shows a slight change in these two ways of estimation. In the third column of Fig.(1), the error percentage we will take in the exact value of $R_c$ and $\lambda_c$ for $R<2$ is $0.1-10\%$ when we ignore the velocity of particles at the final point. On the other hand we see that the critical mass accretion rate increases sharply with $R_f$ for $r<10 r_e$ but at further distances it tends to a constant value, 
\begin{displaymath}
\lambda_c=
\begin{cases}
0.625 &  \text{if}\hspace*{0.3cm}\gamma=1.4\\
0.367 &  \text{if}\hspace*{0.3cm}\gamma=1.6
\end{cases}
\end{displaymath}
In panels (b) and (e) of Fig.(1), it is easily seen that the critical radius decreases notably with moving outwards from the center. Like $\lambda_c$, in the radii about $20 r_e$ we notice that $R_c$ is remained approximately the same,
\begin{displaymath}
R_c=
\begin{cases}
0.2 &  \leftarrow\gamma=1.4\\
0.05  & \leftarrow \gamma=1.6
\end{cases}
\end{displaymath}
Consequently, we find out that neglecting the kinetic energy at the outer radius in Bernoulli function does not change remarkably the  results so we exclude this term in our calculations for the other sections of this paper. 

\section{Spherical accretion in Hernquist galaxy}
As we mentioned before, in this paper we would like to focus on Hernquist galaxy with known gravitational potential and luminosity (Hernquist 1990). We assume that there are two main sources to provide $\Phi$, the first one is the central black hole and the second one is due to the other effective mass of galaxy, so we have
\begin{equation}
\Phi=-\frac{GM_{BH}}{r}-\frac{G M_g}{r+r_g}
\end{equation}
where $M_{BH}$, $M_g$ and $r_g$ are the mass of central blackhole, the total galaxy mass (without $M_{BH}$) and a characteristic scale-length, respectively. Employing dimensionless quantities in Eq.(6) gives,
\begin{displaymath}
\Phi(R)=-c_{sf}^2\bigg(\frac{1}{R}+\frac{m_g}{R+R_g}\bigg)
\end{displaymath}
where $m_g=M_g/M_{BH}$ and $R_g=r_g/r_e$. Here the final form of the function $g(R)$ becomes,
\begin{equation}
g(R)=R^{4\frac{\gamma-1}{\gamma+1}}\bigg(\frac{1}{R}+\frac{m_g}{R+R_g}+\frac{1}{\gamma-1}-\frac{1}{R_f}-\frac{m_g}{R_f+R_g}\bigg)
\end{equation}
We evaluate the extreme value of $g(R)$ with the aim of derivation, 
\begin{displaymath}
\frac{dg(R)}{dR}=0,\hspace*{0.4cm} \rightarrow R=R_c
\end{displaymath}
In this way we find an analytical solution for $R_c$ but too long to write here. We have plotted this new $R_c$ with respect to $R_f$ for different values of $m_g$ and $R_g$ in Fig.(2). According to the panels of this figure, we see $R_c$ decreases rather sharply with respect to (small values of) $R_f$ . Moreover, $R_c$ gradually tends to a constant value in large $R_f$'s. As seen, the gravity of galaxy affects this critical radius too and depending on $R_g$ it can make $R_c$ shorter. The second row panels of this figure shows that the scale-length of galaxy has a direct effect on $R_c$. 

Knowing $R_c$, we can calculate the maximum value of $\lambda$ according to Eq.(18).
According to Fig.(3), $\lambda_c$ experiences a big rise at smaller $R_f$ and then in about $200 R_f$ it maintains approximately the same level. According to the first row panels of Fig.(3), more massive galaxies have larger critical mass accretion rates as we expected. The panels (c) and (d) of Fig.(3) reveals $\lambda_c$ decreases with increasing $R_g$.
\begin{figure*} 
\centering 
\includegraphics[width=190mm]{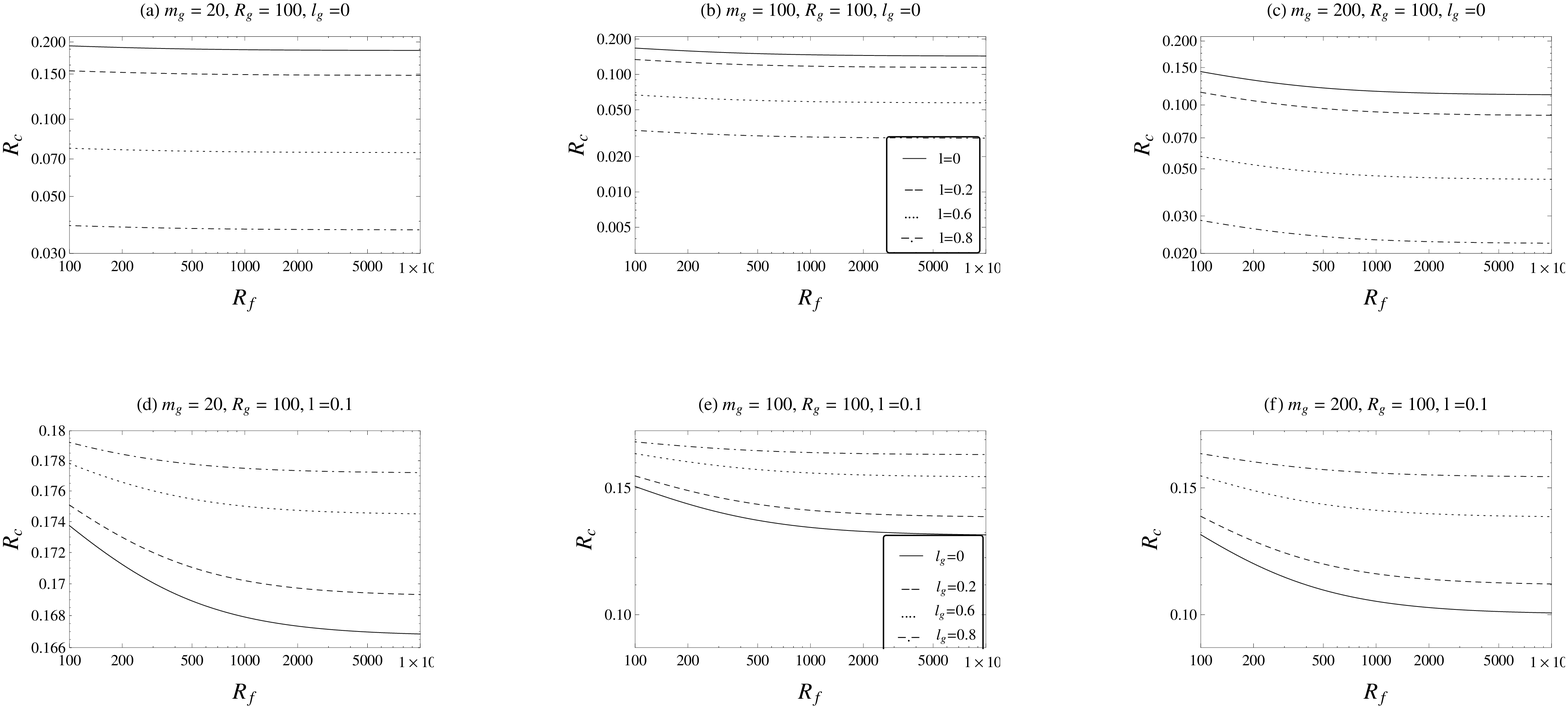} 
\caption{Variation of critical radius, $R_c$ with respect to the final radius of a certain region of accretion, $R_f$. Here we have $\gamma=1.4$ and we have used several values of $l$ and $l_g$ which show the fraction of the luminosity of accretion process to the Eddington limit and the ratio of the luminosity of galaxy to $L_{Edd}$, respectively. In these plots, particles are assumed to have zero velocity at $R_f$.  } 
\end{figure*} 
\section{Adding the luminosity of galaxy in accretion}
In previous section, we have just included the gravity of galaxy in Bernoulli function. Here we add the luminosity of accretion process and also effective brightness of other objects in the galaxy in the momentum equation.  

Now we refer to Eq.(3) and include the extra term of radiation force, $f_{rad}$ as,
\begin{equation}
v_r\frac{dv_r}{dr} = -\frac{1}{\rho}\frac{dp_g}{dr}-\frac{d\Phi}{dr}+\textbf{f}_{rad}
\end{equation}
notice that $f_{rad}$ is a force per unit mass. This force should be specified from its sources. We think of two different factors emeging such a force on the gas particles. The first one is due to the power radiated by the accreting source and it can be of the order of the Eddington Luminosity,
\begin{displaymath}
L=\epsilon L_{Edd}=4\pi GM_{BH}\mu_e m_p\frac{\epsilon}{\sigma}   
\end{displaymath} 
where $c$, $\sigma$ and $\mu_e$ are the light speed, scattering coefficient and the atomic weight per electron, respectively. Following Cassinelli \& Castor (1973), we assume that radiative luminosity of central object, $L$, changes only slightly with radius, so we can take it constant. In the range of low accretion rates we can consider that absorption of photons is negligible and only electron scattering is important, so the effective radiation force exerted on each particles in the gas can be written as  
\begin{displaymath}
{F_{rad}}_1=\frac{\sigma}{c}\frac{L}{4\pi r^2 }
\end{displaymath}
The second source of radiation force that we think about is from the radiation of stars and also thermal radiation of other effective objects in the galaxy but we assume this radiation does not disturb the spherical symmetry of the system. Hence we can define it as, 
\begin{displaymath}
{F_{rad}}_2=\frac{\sigma}{c}\frac{L_g(r)}{4\pi r^2 }
\end{displaymath}
where $L_g(r)$ is the effective luminosity of galaxy at radius $r$. To find $L_g(r)$, we refer to presented model of elliptical galaxies and employ the true 3-D space radiation energy density as (Dehnen 1993,Tremaine et al. 1994),
\begin{equation}
j(r)=\frac{\mathcal{L}}{2\pi}\frac{r_g}{r(r+r_g)^3}
\end{equation} 
where $\mathcal{L}$ is the total luminosity of galaxy (which is obtained if we integrate on $j(r)$ from the center to infinity). We can calculate the luminosity of stars in the galaxy at radius $r$,
\begin{equation}
L_g(r)=\int_0^r 4\pi r^2 j(r) dr=\mathcal{L}\frac{r^2}{(r+r_g)^2},
\end{equation}
Now the total radiation force becomes,
\begin{equation}
F_{rad}=\frac{\sigma}{4\pi c}\big[\frac{L}{r^2}+\frac{\mathcal{L}}{(r+r_g)^2}\big]
\end{equation}
\begin{figure*} 
\centering 
\includegraphics[width=190mm]{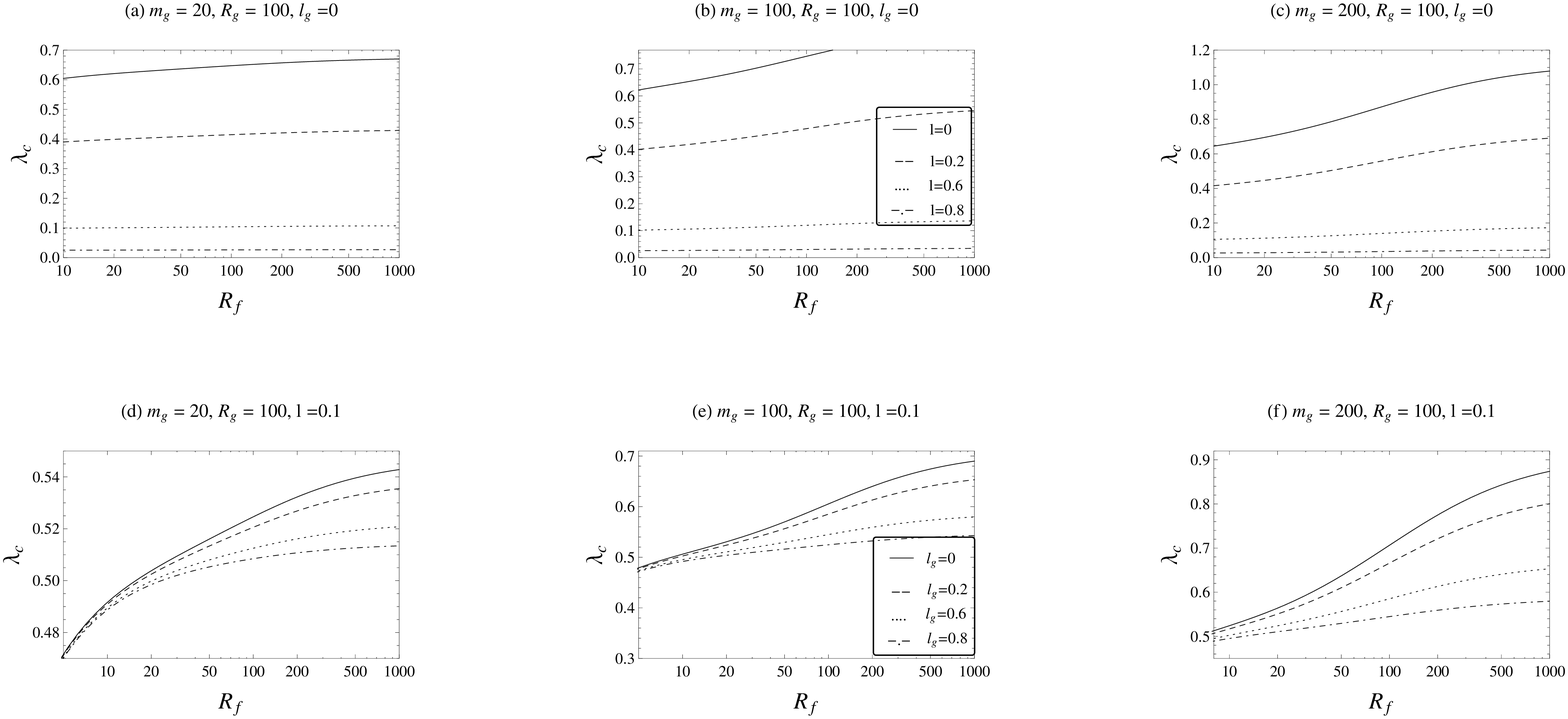} 
\caption{Variation of critical dimensionless mass accretion rate, $\lambda_c$ with respect to the final radius of the selected accretion region, $R_f$. Here  $\gamma=1.4$ and we have examined several values of $l$ and $l_g$. In these plots, particles are assumed to be at rest at the final point. } 
\end{figure*} 
In the following, we scale the luminosities of $L$ and $\mathcal{L}$ with the Eddington luminosity,
\begin{equation}
L_{Edd}=\frac{4\pi GM_{BH}\mu_e m_p c}{\sigma},
\hspace*{0.5cm}
L=l L_{Edd}
\end{equation} 
For scaling $\mathcal{L}$, the formula of the Eddington luminosity becomes,
\begin{equation}
L_{Edd,g}=\frac{4\pi GM_g\mu_e m_p c}{\sigma}=\frac{M_g}{M_{BH}}L_{Edd},
\end{equation}
where we have used Eq.(26) to replace $L_{Edd}$ in it. Consequently we find,
\begin{equation}
\mathcal{L}=l_g L_{Edd,g}=l_g\frac{M_g}{M_{BH}}L_{Edd},
\end{equation}
We substitute Eq.(26) and (28) in Eq. (25) and find the radiation force per unit mass,
\begin{equation}
f_{rad}=\frac{F_{rad}}{\mu_e m_p}=GM_{BH}\big[\frac{l}{r^2}+\frac{M_g}{M_{BH}}\frac{l_g}{(r+r_g)^2}\big],
\end{equation}
The final step here is to integrate of Eq.(29) from $r_f$ to $r$,
\begin{equation}
\int^r_{r_f} f_{rad}=-c_{sf}^2\big[\frac{l}{R}+m_g\frac{l_g}{R+R_g}\big],
\end{equation}
where we have employed dimensionless quantities of Eq.(6) and the galaxy parameters $m_g$ and $r_g$ which are introduced in section 3 of this paper. Integral of Eq.(22) from $r_f$ to the radius $r$ leads us to find,
\begin{displaymath}
\frac{V^2}{2}-\frac{1-l}{R}-\frac{m_g(1-l_g)}{R+R_g}+\frac{C_s^2}{\gamma-1}=
\end{displaymath}
\begin{equation}
\frac{V_f^2}{2}-\frac{1-l}{R_f}-\frac{m_g(1-l_g)}{R_f+R_g}+\frac{1}{\gamma-1}
\end{equation}
where we have substituted Eq.(20) and (4) in it and also using relations of Eq.(6) to make the final equation dimensionless. Now we can specify the function of $g(R)$ as,
\begin{displaymath}
g(R)=R^{4\frac{\gamma-1}{\gamma+1}}\bigg[\frac{(1-l)}{R}+\frac{m_g(1-l_g)}{R+R_g}
\end{displaymath}
\begin{equation}
\hspace*{1cm}-\frac{(1-l)}{R_f}-\frac{m_g(1-l_g)}{R_f+R_g}+\frac{V_f^2}{2}+\frac{1}{\gamma-1}\bigg]
\end{equation}
notice that for plotting graphs, we will ignore the term of $V_f^2/2$ in $g(R)$ because it depends on $\lambda$ and also as we mentioned before it causes a  little change on $R_c$ and $\lambda_c$. 

 In figures (3) and (4), we see the effects of both sources of luminosity in the galaxy on the critical radius and mass accretion rate. In figure (4), we have limited the final radius with this condition $R_f\geq R_g$. According to the panels (a)-(c) of Fig.(4), when there is no source of luminosity, the critical radius of $R_c$ in thin and slim galaxies (i.e. $m_g$ is at least one order smaller than $R_g$) is approximately constant for almost all chosen final radius of accretion region. Comparison with slim galaxy, different values of $R_f$ have more influence on $R_c$ in accretion of dark denser galaxies. On the other hand, if we compare the upper panels of Fig.(4) with the panels in the bottom of it, we see that the influences of $l$ and $l_g$ on $R_c$ are completely different.  
The result of the luminosity from the first source in panels (a)-(c) is expectable, but the second luminosity on $R_c$ acts  in the opposite direction and makes the critical point further from the center. Panels (d-f) say that $R_c$ expands with the luminosity of galaxy. In figure (5), $\lambda_c$ presents similar trend with changing both $l$ and $l_g$ and decreases by increasing them. 

In figure (6), we have assumed $R_f$ tends to infinity and studied the variation of $\lambda_c$ and $R_c$ with respect to the effective linear density of Hernquist galaxy, i.e. $m_g/R_g$ with several sets of $l$ and $l_g$. According to panel (a) of Fig.(6), we find out that there is a linear dependence between $\lambda_c$ and $m_g/R_g$. Moreover, the effect of the luminosity from accretion is stronger than the luminosity of galaxy itself and decreases much more $\lambda_c$. The second panel shows that $R_c$ experiences significantly decreases in thiner dark galaxies. The inverse effects of $l$ and $l_g$ on the critical radius are easily seen in this plot too.     
\begin{figure*} 
\centering 
\includegraphics[width=190mm]{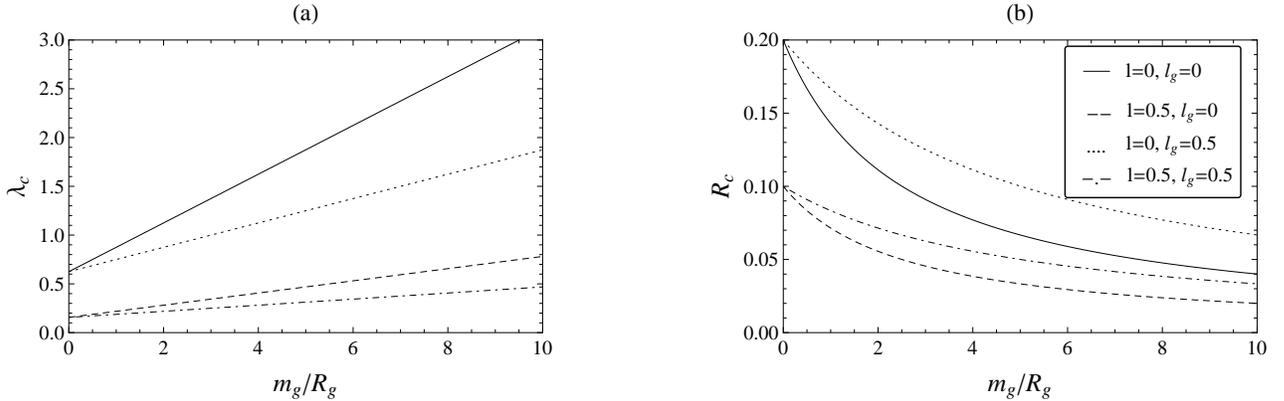} 
\caption{Variation of critical accretion parameter, $\lambda_c$ and critical radius, $R_c$ with respect to the characteristic dimensionless linear density of galaxy, i.e. $m_g/R_g$. Here,  $\gamma=1.4$ and we have examined several values of $l$ and $l_g$. Following Bondi's model, We have assumed that both the kinetic and potential energies of particles are zero at the end of accretion region.  } 
\end{figure*} 
\section{Transonic solutions' Type}
Up to now, we have examined the positions of critical points and obtained several relations on accretion rates. Here, we are able to study the nature of critical points. According to previous studies (Kato et al. 2008), there are generally four types of critical points (topologies): saddle, center, node and spiral. In the present case that we have an adiabatic flow without viscosity and other dissipative processes, we expect that the transonic point to be saddle. To be assured that our expectation is satisfied, we refer to equations (13), (16) and (32),
\begin{displaymath}
\big(\frac{u^2}{2}+\frac{1}{\gamma-1}\big)\big(\frac{\lambda}{R^2 u}\big)^{2\frac{\gamma-1}{\gamma+1}}=\frac{(1-l)}{R}+\frac{m_g(1-l_g)}{R+R_g}
\end{displaymath}
\begin{equation}
\hspace*{0.5cm}-\frac{(1-l)}{R_f}-\frac{m_g(1-l_g)}{R_f+R_g}+\frac{1}{\gamma-1}
\end{equation}
We can solve this equation with given parameters and at each point with a certain radius. In figures (7) and (8), we have shown the variation of Mach number, $u$, with respect to the dimensionless radius, $R$. Fig.(7) is devoted to the effect of accretion rate on $u$. As seen, there are two solutions for $u$ per each $\lambda<\lambda_c$. In the critical point, $R_c$ with $\lambda=\lambda_c$, there is just one solution for Eq.(33), and Fig.(7) clearly shows the slop of $u$'s curve completely changes and becomes positive at $R=R_c$. According to the four panels of Fig.(8), we see other values of $m_g$ and $R_g$ or adding luminosity of two sources in $g(R)$ do not disrupt this trend of Mach number at the vicinity of critical points. Consequently, the type of transonic solutions in the present problem is saddle.
\begin{figure} 
\centering 
\includegraphics[width=80mm]{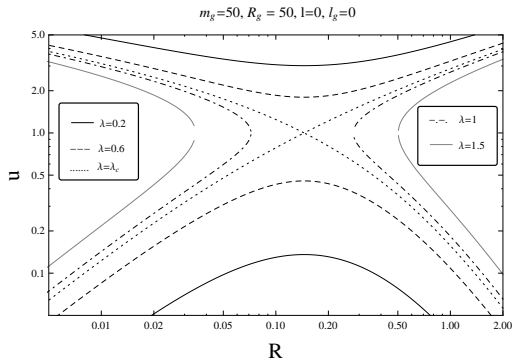} 
\caption{Mach number, $u$, as a function of the dimensionless radius, $R$.  Here,  $\gamma=1.4$ and we have examined several  $\lambda$'s. One of them is equal to the critical value of accretin parameter, $\lambda_c=0.8621$.    } 
\end{figure}
\begin{figure*} 
\centering 
\includegraphics[width=190mm]{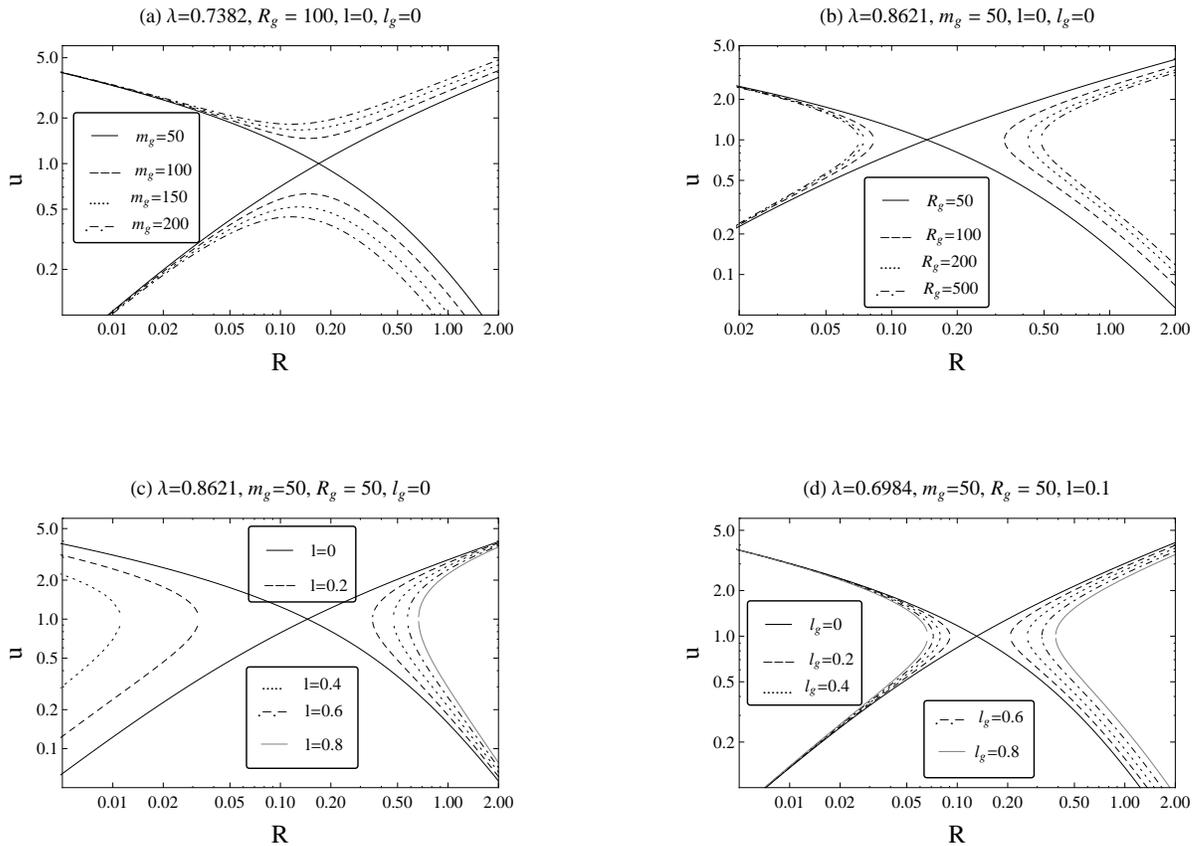} 
\caption{Mach number, $u$, as a function of the dimensionless radius, $R$.  Here,  $\gamma=1.4$ and we can see how changing each parameter in the potential energy affect Mach number. } 
\end{figure*}

\section{Summary and Conclusion}
In this paper, we reconsidered the basic and fundamental model of accretion which was studied by Bondi (1952) and followed by many authors after him. We wrote the basic equations of hydrodynamics and simplified them by these assumptions 1)steady state, 2)spherical symmetry and 3) polytropic relation between pressure and density of gas. To find the constant value of Bernoulli function, we used a limited radius for integral on the momentum equation (in Bondi's paper one limit of integral was infinity). Hence, we had three constant terms from the total energy of this system at the final radius ($r_f$) of the selected accretion region. We found out that kinetic energy at $r_f$ was ignorable for $r_f/(GM/c_{sf}^2)>2$, thus we just included the gravity potential besides thermal energy at the final point to estimate the value of Bernoulli function. We employed the spherical accretion model for Hernquist galaxies and substituted the gravity potential with the known gravity of Hernquist model. Two factors were added in the mentioned model, the first one is $m_g$ which is the ratio of the galaxy's total mass ($M_g$) to the central blackhole's mass ($M_{BH}$), and the second one is $R_g$ is the fraction of the characteristic scale-length of galaxy ($r_g$) to the effective length of accretion ($r_e=GM_{BH}/c_{sf}^2$), respectively. We noticed that the maximum possible accretion rate ($\lambda_c$) was influenced remarkably by these two factors and behaved in two different trends, means $m_g$ increased the accretion rate but $R_g$ decreased it. Moreover, the value of $\lambda_c$ became independent of $r_f$ as long as it is about 3 order of magnitude larger than $GM_{BH}/c_{sf}^2$. The other factor that we studied in this paper was the luminosity from two sources, one $L$ produced by the accretion process , and the other one $\mathcal{L}$ from the galaxy itself. According to the Hernquist model, we could evaluate the radiation force ($F_{rad}$) on particles but we just included electron scattering in the opacity of gas. Therefore, $F_{rad}$ converted to a simple form and with the help of Eddington luminosity definition became a term like gravity force, one term proportional to the squared radius $\propto r^{-2}$ and the second term $\propto (r+r_g)^{-2}$. To parameterize $L$ and $\mathcal{L}$, we compared them with the Eddington luminosity $L_{Edd}$ and introduced two parameter $l(=L/L_{Edd})$ and $l_g(=L/L_{Edd})$. These two parameters had opposite effect on the critical radius of accretion $R_c$ (where the velocity becomes equal to the sound speed) and it decreases with increasing $l$ whereas $l_g$ made $R_c$ larger. However, both $l$ and $l_g$ had a negative influence on $\lambda_c$ and decreased it. 

For future studies, we will not restrict ourself to consider optically thin mediums and include the absorption opacity of gas for finding the effect of radiation force on accreting particles and find more general solutions. The other work is to study the possible changes of self-gravity of galaxy on the maximum mass accretion rate.

\section*{Acknowledgements}
This work has been supported financially by Research Institute for
Astronomy and Astrophysics of Maragha (RIAAM) under research
project No. 1/6275-1.


\begin{thebibliography}{99} 
\bibitem[]{}Allen, S. W., Dunn, R. J. H., Fabian, A. C., Taylor, G. B., Reynolds, C. S., 2006, MNRAS, 372, 21
\bibitem[]{}Barai, P., Proga, D., \& Nagamine, K. 2011, MNRAS, 418, 591
\bibitem[]{}Barai, P., Proga, D., \& Nagamine, K. 2012, MNRAS, 424, 728
\bibitem[]{}Beckmann, R. S., Slyz, A., Devriendt, J. 2018, MNRAS 478, 995 
\bibitem[]{}Begelman M. C., 1978, MNRAS, 184, 53
\bibitem[]{}Begelman M. C., 1979, MNRAS, 187, 237
\bibitem[]{}Bondi H., 1952, MNRAS, 112, 195 
\bibitem[]{} Ciotti L., \& Pellegrini S. 2017, ApJ, 848, 29
\bibitem[]{}Ciotti L. \& Pellegrini S., 2018, ApJ, 868, 91 
\bibitem[]{}Dehnen W., 1993, MNRAS, 265, 250
\bibitem[]{}Edgar R., 2004, New Astronomy Reviews, 48, 843
\bibitem[]{}Fukue J., 2001, PASJ, 53, 687
\bibitem[]{}Gallo, E., et al. 2010, ApJ 714, 25 
\bibitem[]{}Hernquist, L., 1990, ApJ, 356, 359 
\bibitem[]{}Kato S., Fukue J., Mineshige S., 2008, Black-Hole Accretion Disks –
Towards a New Paradigm. Kyoto Univ. Press, Kyoto 
\bibitem[]{}Korol V., Ciotti L., \& Pellegrini S. 2016, MNRAS, 460, 1188 
\bibitem[]{}Kaaz N., Antoni A., Ramirez-Ruiz E., 2019, ApJ, 876, 142
\bibitem[]{}Loewenstein, M., Mushotzky, R. F., Angelini, L., Arnaud, K. A., Quataert, Eliot 2001, ApJ, 555, 21 
\bibitem[]{}Maraschi L., Reina C., \& Treves A., 1974, A\& A, 35, 389
\bibitem[]{}Mathews W. G., \& Guo F., 2012, ApJ, 754, 154
\bibitem[]{}McNamara, B. R., Rohanizadegan, M., Nulsen, P. E. J. 2011, ApJ, 727, 39
\bibitem[]{}Park, K.-H., Wise, J. H., \& Bogdanovi ć, T. 2017, ApJ, 847, 70
\bibitem[]{}Pellegrini, S. 2005, ApJ, 624, 155
\bibitem[]{}Pellegrini, S. 2010, ApJ, 717, 640 
\bibitem[]{}Ramirez-Velasquez J. M., Sigalotti L., Gabbasov R., Cruz F., Klapp J., Contreras E., 2019, arXiv:1902.00360
\bibitem[]{}Ruffert M.; Melia F., 1994, A\& A, 228L, L29
\bibitem[]{}Russell, H. R., Fabian, A. C., McNamara, B., Broderick, A. E., 2015, MNRAS, 451, 588
\bibitem[]{}Sijacki, D., Springel, V., Di Matteo, T., \& Hernquist, L. 2007, MNRAS, 380, 877 
\bibitem[]{}Tremaine S., Richstone D. O., Byun Y.-I., Dressler A., Faber S. M., Grillmair C., Kormendy J., Lauer T. R., 1994, AJ, 107, 634 
\bibitem[]{}Wong, Ka-Wah, Irwin, J.A., Shcherbakov, R. V., Yukita, M., Million, E. T., Bregman, J. N. 2014, ApJ, 780, 9
\bibitem[]{}Yalinewich A., Sari R., Generozov A., Stone N. C.; Metzger B. D., 2018, MNRAS, 479, 4778
\end{thebibliography}
\end{document}